\documentclass[10pt, a4paper, twocolumn]{article} 
\usepackage[english]{babel} 

\usepackage{microtype} 

\usepackage{amsmath,amsfonts,amsthm} 

\usepackage[svgnames]{xcolor} 

\usepackage[hang, small, labelfont=bf, up, textfont=it]{caption} 

\usepackage{booktabs} 

\usepackage{lastpage} 

\usepackage{graphicx} 

\usepackage{enumitem} 

\usepackage{listings}

\usepackage{colortbl}

\usepackage[euler]{textgreek}

\setlist{noitemsep} 

\usepackage{sectsty} 
\allsectionsfont{\usefont{OT1}{phv}{b}{n}} 

\usepackage{geometry} 

\usepackage[T1]{fontenc} 
\usepackage[utf8]{inputenc} 

\usepackage{XCharter} 

\usepackage{fancyhdr} 
\fancypagestyle{firstpage}{ 
	\fancyhf{}
}

\newcommand{\authorstyle}[1]{{\large\usefont{OT1}{phv}{b}{n}\color{DarkRed}#1}} 

\newcommand{\institution}[1]{{\footnotesize\usefont{OT1}{phv}{m}{sl}\color{Black}#1}} 

\usepackage{titling} 

\newcommand{\HorRule}{\color{DarkGoldenrod}\rule{\linewidth}{1pt}} 

\pretitle{
	\vspace{-30pt} 
	\HorRule\vspace{10pt} 
	\fontsize{32}{36}\usefont{OT1}{phv}{b}{n}\selectfont 
	\color{DarkRed} 
}

\posttitle{\par\vskip 15pt} 

\preauthor{} 

\postauthor{ 
	\vspace{10pt} 
	\par\HorRule 
	\vspace{20pt} 
}

\usepackage{lettrine} 
\usepackage{fix-cm}	

\newcommand{\initial}[1]{ 
	\lettrine[lines=3,findent=4pt,nindent=0pt]{
		\color{DarkGoldenrod}
		{#1}
	}{}%
}

\usepackage{xstring} 

\newcommand{\lettrineabstract}[1]{
	\StrLeft{#1}{1}[\firstletter] 
	\initial{\firstletter}\textbf{\StrGobbleLeft{#1}{1}} 
}

\usepackage[backend=bibtex,sorting=none,citestyle=numeric,natbib=true]{biblatex} 

\usepackage[autostyle=true]{csquotes} 

\title{Influence of atomic FAA on ParallelFor and a cost model for improvements}
\author{
  \authorstyle{Ran Shuai}\\
  \institution{Microsoft, Silicon Valley Campus, Mountain View, CA, United States}\\
  \institution{\texttt{rashuai@microsoft.com}}
}
\date{\today}
\begin{document}
\maketitle
\thispagestyle{firstpage}
\lettrineabstract{Abstract - this paper focuses on one of the most frequently visited multithreading library interfaces - ParallelFor. In this study, it is inferred that ParallelFor’s end-to-end latency performance is noticeably affected by the frequency with which fetch-and-add (FAA) is called during program execution. This can be explained by ParallelFor’s uniform semantics and the utilization of atomic FAA. To prove this assumption, a battery of tests was designed and conducted on diverse platforms. From the collected performance statistics and overall trends, several conclusions were drawn and a cost model is proposed to enhance performance by mitigating the influence of FAA.}

\section*{Introduction}
For decades, increasingly high-performance workstations have been adopted in production environments to boost productivity. These machines are commonly equipped with multi-core CPU(s) that allow for parallelism in applications and services. Each CPU core hosts one or two hardware threads (e.g., hyper-threading) that undertake assigned tasks independently. With a suitable task-breakdown policy and wait-and-join semantics, an application or service could allow several of its tasks to run simultaneously, instead of sequentially, which reduces latency and promotes responsiveness.\cite{Marr2002HyperThreadingTA} \cite{Roberts2006MultiCorePI} \cite{1237999} 

To fully utilize the hardware’s computing power, multi-threading software libraries serve to bridge upstream software applications and hardware threads. Usually, such libraries create, host, and maintain a set of software threads known as a thread pool. Moreover, the libraries provide API(s), such as ParallelFor(…) or ForEach(…), that take a task/function and an integer number as the iterations to run. Upon being called, the libraries will distribute the task to all threads in the pool. When the maximum iterations are reached and all threads have completed the assigned iterations, ParallelFor(…) or ForEach(…) will return to the caller. One such scenario is openmp \cite{660313}---a built-in library for C++ compilers, such as GNU gcc \cite{GCCINTRO} and LLVM \cite{LLVM}. intelTBB and a recently published library---TaskFlow \cite{9511796} serve similar purposes.

According to Amdahl's law \cite{4563876}, a parallelized application could provide end-to-end performance benefits as:

\[SpeedUp(T) = 1 / ((1-P) + P/T)\]

where T is the number of threads and P is the parallelizable fraction of the application. For ParallelFor, P is always 1; thus, we could expect the gain to be:

\[SpeedUp(T) = T\]

\section*{Problem statement}
The measurement of performance gains should neither be limited to Amdahl's law, nor the assumption that n threads in total ``should'' bring down the overall cost to total\_cost/number\_of\_threads. The parallelism provided by multit-hreading libraries leads to some nontrivial costs that could challenge the estimation of end-to-end latency. Considering ParallelFor as an example, its most representative implementations follow a simple semantic:

\begin{lstlisting}[language=C++]
void ParallelFor(
  function<void(int)>& task,
  int N) {

  atomic<int> counter{0};
  int block_size{...};
  function<void(void)> thread_task =
    [&] () {
      int begin{0};
      while ((begin = 
        counter.FetchAndAdd(block_size))<N){
        for (int iter = begin;
             iter < min(N,begin+block_size);
             iter++) {
          task(iter);
        }
      }
  };
  for(Thread& thread: threadpool) {
    thread.Enqueue(thread_task);
  }
  thread_task();
  for (Thread& thread: threadpool) {
    thread.wait(...)
  }
}
\end{lstlisting}

As seen in this code snippet, the ParallelFor function first wraps up the input task with a thread task, which utilizes atomic fetch-and-add (FAA) to acquire a range of iterations for execution. The atomic component serves to synchronize between threads to avoid a data race. Then, ParallelFor assigns the thread task to all threads in the pool and waits for all the returns. Based on the listed logic, the ParallelFor’s caller could be assured that the input task will be called exactly N times, with an input iterating from 0 to N-1.

Among all the listed details, ``block\_size'' is a key variable that may significantly affect the overall end-to-end latency. It decides for how many iterations the thread should run each input task in the ``for'' loop. Equivalently, ``block\_size'' is a variant that determines the size by which ``N'' could be divided.

The issue here is that the calling of atomic FAA results in nontrivial cost because cache invalidation and resynchronization are required to ensure that all threads visiting the atomic variable see the same value. \cite{schweizer2020evaluating} presented a keen estimation on atomic overheads as follows:

\[L(A, S) = R(S) + E(A) + O\]

where A is the type of atomic operation, such as FAA, compare-and-swap (CAS), or swap (SWP); S is the cache state among {Modified, Owned, Exclusive, Shared, Invalid} \cite{MOESI}, R(S) is the time-cost of acquiring ownership of the cache in S state, E(A) is the cost of performing A on the cache, and O represents other miscellaneous costs. In our case, A is always FAA and S is assumed to be Shared; hence, a simplified version of our case is:

\[L = R + E + O\]

According to \cite{schweizer2020evaluating}, R takes the most significant part of L, implying that threads spent most of the time during the interval on acquiring ownership of the cache. Furthermore, the number of threads and their locality also influence the cost of R, and therefore, the overall L. For example, two threads in the cores sharing the same L3 cache tend to spend less time on R than when the threads are in cores across sockets.

Assuming that the total number of iterations N is to be divided into several blocks of block size B, and the number of threads is T. Consequently, the overall end-to-end latency of ParallelFor may be formulated as

\[Cost(T, N, L) = N/B*L + O(N)/T\]

Based on this formula, we could reasonably infer that Cost(T, N, L) is lower when B is larger, i.e., a larger block size means less calling of FAA, resulting in a lower total atomic overhead.

But what is the upper bound of B? We cannot assert that Cost(T, N, L) will always decrease as the block sizes increase. For instance, considering that we have T threads fully at our disposal, by setting B to exceed N/T, only fewer than T threads are allowed the chance to run the input task---the potential of parallelism is thereby jeopardized.
The test that follows shows that the end-to-end latency would begin falling before B reaches N/T because the threads are scheduled to run on a physical CPU core at different times, and as the load sharing among cores is occasionally imbalanced, the threads running time may vary for a period, which implies that smaller sized blocks are more likely to match the threads' running quota perfectly.
\cite{rajagopalan2007thread}.

Hence, knowing that the block size is expected to be larger to avoid excessive atomic FAA overhead, it should not exceed a certain limit. The problem lies in how to determine the proper B value.

\section*{Test and statistics}

To address this problem, a series of tests was conducted to resolve the problem of determining how FAA with varied block sizes would affect the end-to-end latency of ParallelFor. Additionally, different ``sized'' tasks were included, as a task that reads and writes 32 bytes of memory with simple computations would not have the same performance as a more complex task. Furthermore, a task that works mostly on IO should behave differently from one that is more CPU intensive. To address these variants, we implemented a configurable unit task function:
\begin{lstlisting}[language=C++]
unit_task =
  [unit_read, unit_write,
  unit_computation, ...] (...) {

  uint8_t* read_at =  ...;
  uint8_t* write_at = ...;
  uint64_t per_read_computation =
      unit_computation / unit_read;
  uint64_t write_count = 0;
  uint8_t integer = 0;

  for (uint64_t i = 0; i < unit_read; ++i) {
      integer = read_at[i];
      for (uint64_t j = 0; j < 
           per_read_computation; ++j) {
          integer += 1;
      }
      if (write_count < unit_write) {
          write_at[write_count++] = integer;
      }
  }
  while (write_count < unit_write) {
      write_at[write_count++] = integer;
  }
};
\end{lstlisting}

Note that unit\_task\ will be sent to ParallelFor as an input argument. As observed from the implementation, unit\_task\ references three external variables \- unit\_read, unit\_write, and unit\_computation. Here, unit\_read denotes the number of bytes of memory to read; unit\_write the number of bytes to write; and unit\_computation the number of computations that should occur along with execution. By using these variables, we could run tests with tasks of varied ``sizes'' to obtain unbiased conclusions.

Furthermore, we implemented a thread pool to provide the ParallelFor function, following the semantic listed in the previous section. The thread pool allows a number of threads with fixed affinity settings to be configured to restrict the threads on certain cores. This helps to reduce the noise from thread rescheduling between cores and to maintain load balancing.

Finally, to ensure that the test results are more generalized, we prepared a set of platforms with diverse hardware specifications and computing capabilities.

\begin{itemize}
\item Dell M4 workstation with Intel Xeon® W-3225R @ 3.70 GHZ, Windows 10 Pro 18362.1171
\item AMD Ryzen Threadripper 3970X 32-Core Processor @ 3.69 GHz, Windows 10 Pro 19042.1165
\item Dell M4 workstation with two Intel® Xeon® Gold 5225R @ 2.20 GHZ, Windows 10 Pro 19042.1165
\end{itemize}

The W-3225R has one CPU of eight cores, each has its proprietary L1 and L2 but share the same L3; the AMD 3970X has one CPU of 32 cores, every 4 cores share an L3; the Gold 5225R has 2 CPUs sitting on separate sockets, each has 24 cores sharing the same L3, whereas L1 and L2 are core private. By using hwloc \cite{hwloc}, graphs of the internal hardware topology were created as follows:

\begingroup\centering

\includegraphics[width=8.1cm, height=7cm]{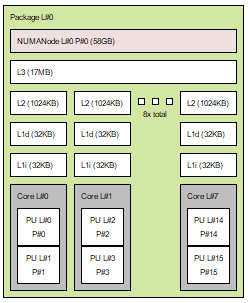}
\textit{Intel W-3225R}

\includegraphics[width=8cm, height=15cm]{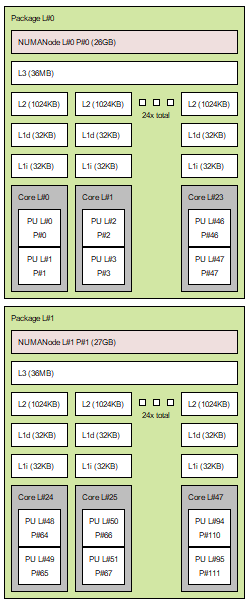}
\textit{Intel Gold 5225R}

\includegraphics[width=8cm, height=24.2cm]{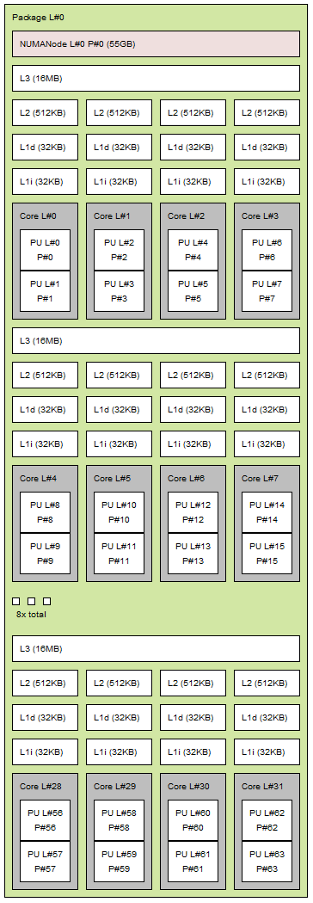}
\textit{AMD Ryzen Threadripper 3970X}

\endgroup

\bigbreak
With the task, thread pool, and platforms all implemented and ready, the end-to-end ParallelFor performance was first tested with different block sizes, varying unit computation, and thread pool sizes on the W-3225R. The latency was measured using the CPU clocks: 
\smallbreak

\begingroup\centering

\begin{tabular}{|p{1.6cm}|p{1.6cm}|p{1.6cm}|p{1.6cm}|}
  \hline
  \multicolumn{4}{|c|}{unit read 1024, unit write 1024, unit comp 1024} \\
  \hline
  block sizes&2 threads&4 threads&8 threads\\
  \hline
  1&1394900&957700&569100\\
  2&1291900&762200&\cellcolor{LightGreen}445600\\
  4&1240400&706200&\cellcolor{LightGreen}430200\\
  8&1112800&\cellcolor{LightGreen}644700&\cellcolor{LightGreen}462000\\
  16&\cellcolor{LightGreen}1078100&\cellcolor{LightGreen}646900&\cellcolor{LightGreen}437200\\
  32&\cellcolor{LightGreen}1060400&\cellcolor{LightGreen}643700&\cellcolor{LightGreen}447000\\
  64&\cellcolor{LightGreen}1082100&\cellcolor{LightGreen}643000&494000\\
  128&\cellcolor{LightGreen}1081400&691700&558000\\
  256&1192000&798200&632000\\
  512&1179800&796900&800500\\
  1024&1585000&799100&1309500\\
  \hline
\end{tabular}
\smallbreak
\textit{W-3225R: unit comp 1024, e2e latency in clocks}
\smallbreak

\begin{tabular}{ |p{1.6cm}|p{1.6cm}|p{1.6cm}|p{1.6cm}| }
  \hline
  \multicolumn{4}{|c|}{unit read 1024, unit write 1024, unit comp \begin{math}1024^3\end{math}} \\
  \hline
  block sizes&2 threads&4 threads&8 threads\\
  \hline
  1&1159900&859900&465500\\
  2&1062200&755600&\cellcolor{LightGreen}419200\\
  4&1048400&692200&\cellcolor{LightGreen}4132000\\
  8&1028100&698500&\cellcolor{LightGreen}416200\\
  16&\cellcolor{LightGreen}979000&\cellcolor{LightGreen}629200&\cellcolor{LightGreen}418100\\
  32&\cellcolor{LightGreen}978500&\cellcolor{LightGreen}633400&450500\\
  64&997400&640300&474900\\
  128&985900&691900&571100\\
  256&1038200&698100&549700\\
  512&1192700&785600&788800\\
  1024&1386200&793300&819000\\
  \hline
\end{tabular}
\smallbreak
\textit{W-3225R: unit comp \begin{math}1024^3\end{math}, e2e latency in clocks}
\smallbreak

\begin{tabular}{ |p{1.6cm}|p{1.6cm}|p{1.6cm}|p{1.6cm}| }
  \hline
  \multicolumn{4}{|c|}{unit read 1024, unit write 1024, unit comp \begin{math}1024^4\end{math}} \\
  \hline
  block sizes&2 threads&4 threads&8 threads\\
  \hline
  1&1307100&689200&430600\\
  2&1236200&638500&392900\\
  4&1203300&\cellcolor{LightGreen}604200&\cellcolor{LightGreen}382900\\
  8&\cellcolor{LightGreen}1197900&\cellcolor{LightGreen}594800&\cellcolor{LightGreen}380300\\
  16&\cellcolor{LightGreen}1159000&\cellcolor{LightGreen}607000&\cellcolor{LightGreen}374800\\
  32&\cellcolor{LightGreen}1197300&\cellcolor{LightGreen}594500&389800\\
  64&1232400&643700&431800\\
  128&1274800&609400&487100\\
  256&1275100&698000&589800\\
  512&1233500&785300&691400\\
  1024&1402400&1178600&1091800\\
  \hline
\end{tabular}
\smallbreak
\textit{W-3225R: unit comp \begin{math}1024^4\end{math}, e2e latency in clocks}
\smallbreak

\endgroup

For the listed cases, the best performance varies by block size. However, noting the highlighted cells, it is apparent that mounting the unit computation decreases the most preferred block size, and increasing the number of threads that participate in computation produces a similar outcome. To confirm the observation, we conducted the following tests on the G-5225R:
\smallbreak

\begingroup\centering

\begin{tabular}{ |p{1.6cm}|p{1.6cm}|p{1.6cm}|p{1.6cm}| }
  \hline
  \multicolumn{4}{|c|}{unit read 1024, unit write 1024, unit comp \begin{math}1024^3\end{math}} \\
  \hline
  block sizes&4 threads&8 threads&16 threads\\
  \hline
  1&948300&555200&311700\\
  2&900800&532700&\cellcolor{LightGreen}272200\\
  4&\cellcolor{LightGreen}872500&\cellcolor{LightGreen}516900&\cellcolor{LightGreen}267300\\
  8&\cellcolor{LightGreen}868900&\cellcolor{LightGreen}522600&\cellcolor{LightGreen}269100\\
  16&\cellcolor{LightGreen}865200&\cellcolor{LightGreen}505900&\cellcolor{LightGreen}283100\\
  32&\cellcolor{LightGreen}864000&\cellcolor{LightGreen}512300&302600\\
  64&\cellcolor{LightGreen}889200&\cellcolor{LightGreen}502500&366500\\
  128&\cellcolor{LightGreen}874500&\cellcolor{LightGreen}512200&341600\\
  256&1023300&665100&511700\\
  512&1041200&809700&1003300\\
  1024&1332100&1334300&1584000\\
  \hline
\end{tabular}
\smallbreak
\textit{Gold 5225R: \begin{math}1024^3\end{math} unit comp}
\medbreak

\begin{tabular}{ |p{1.6cm}|p{1.6cm}|p{1.6cm}|p{1.6cm}| }
  \hline
  \multicolumn{4}{|c|}{unit read 1024, unit write 1024, unit comp \begin{math}1024^5\end{math}} \\
  \hline
  block sizes&4 threads&8 threads&16 threads\\
  \hline
  1&889300&545300&314700\\
  2&847100&\cellcolor{LightGreen}494100&\cellcolor{LightGreen}269500\\
  4&\cellcolor{LightGreen}826500&\cellcolor{LightGreen}495900&\cellcolor{LightGreen}263700\\
  8&\cellcolor{LightGreen}822900&\cellcolor{LightGreen}484400&\cellcolor{LightGreen}288100\\
  16&\cellcolor{LightGreen}827500&\cellcolor{LightGreen}489800&\cellcolor{LightGreen}273500\\
  32&\cellcolor{LightGreen}839300&\cellcolor{LightGreen}498900&296000\\
  64&\cellcolor{LightGreen}817400&513000&305600\\
  128&898800&502500&336600\\
  256&1025700&515600&389600\\
  512&1009900&661400&674600\\
  1024&1015400&1335000&1037500\\
  \hline
\end{tabular}
\smallbreak
\textit{Gold 5225R: \begin{math}1024^5\end{math} unit comp}
\medbreak

\begin{tabular}{ |p{1.6cm}|p{1.6cm}|p{1.6cm}|p{1.6cm}| }
  \hline
  \multicolumn{4}{|c|}{unit read 1024, unit write 1024, unit comp \begin{math}1024^6\end{math}} \\
  \hline
  block sizes&4 threads&8 threads&16 threads\\
  \hline
  1&913000&545000&313100\\
  2&860600&521200&281000\\
  4&830100&496800&\cellcolor{LightGreen}263300\\
  8&832000&503200&\cellcolor{LightGreen}264500\\
  16&827500&\cellcolor{LightGreen}498200&273400\\
  32&\cellcolor{LightGreen}814300&512300&303400\\
  64&827200&502000&294700\\
  128&899600&517700&337200\\
  256&1013100&671700&412000\\
  512&1036000&660700&673700\\
  1024&1013900&1337500&1022000\\
  \hline
\end{tabular}
\smallbreak
\textit{Gold 5225R: \begin{math}1024^6\end{math} unit comp}
\medbreak

\endgroup

The Gold 5225R cases confirm that a larger unit computation reduces the preferred block size. That there is a constant range of how much computation a thread can handle in its average CPU quota is a possible explanation. This means that the larger the unit computation, the smaller the block size should be to keep the overall block computation in that range. Note: 

\[computation\_of\_block = block\_size * unit\_comp\]

In addition, FAA overheads are low because threads in the tests are running on cores sharing the same L3, thus by adding more threads, block size should be kept smaller to allow for better parallelism. Contrastingly, by defining cores that share the same L3 as a core group, the opposite trend was observed when adding more core groups to the tests:

\smallbreak
\begingroup\centering

\begin{tabular}{ |p{1.6cm}|p{1.6cm}|p{1.6cm}|p{1.6cm}| }
  \hline
  \multicolumn{4}{|c|}{unit read 1024, unit write 1024, unit comp \begin{math}1024^2\end{math}} \\
  \hline
  block sizes&24 threads&36 threads&48 threads\\
  \hline
  1&309600&325100&490600\\
  2&269400&234400&498600\\
  4&302000&228600&381200\\
  8&\cellcolor{LightGreen}264700&274800&376700\\
  16&\cellcolor{LightGreen}273700&235500&236100\\
  32&296400&\cellcolor{LightGreen}257900&\cellcolor{LightGreen}212000\\
  64&301900&\cellcolor{LightGreen}257800&\cellcolor{LightGreen}193600\\
  128&337000&360400&420400\\
  256&389100&406000&553600\\
  512&701000&840800&1161600\\
  1024&1066900&1744600&2402800\\
  \hline
\end{tabular}
\smallbreak
\textit{Gold 5225R: \begin{math}1024^2\end{math} unit comp}
\medbreak

\begin{tabular}{ |p{1.6cm}|p{1.6cm}|p{1.6cm}|p{1.6cm}| }
  \hline
  \multicolumn{4}{|c|}{unit read 1024, unit write 1024, unit comp \begin{math}1024^4\end{math}} \\
  \hline
  block sizes&24 threads&36 threads&48 threads\\
  \hline
  1&280000&552400&588200\\
  2&\cellcolor{LightGreen}238800&538000&514800\\
  4&\cellcolor{LightGreen}232700&475300&475300\\
  8&268900&307600&358900\\
  16&235600&\cellcolor{LightGreen}226400&\cellcolor{LightGreen}279100\\
  32&259700&\cellcolor{LightGreen}245400&360500\\
  64&259400&\cellcolor{LightGreen}290900&\cellcolor{LightGreen}287600\\
  128&350500&366900&282600\\
  256&436300&610400&496100\\
  512&816500&1221700&868500\\
  1024&1686600&2222300&1472500\\
  \hline
\end{tabular}
\smallbreak
\textit{Gold 5225R: \begin{math}1024^4\end{math} unit comp}
\medbreak

\begin{tabular}{ |p{1.6cm}|p{1.6cm}|p{1.6cm}|p{1.6cm}| }
  \hline
  \multicolumn{4}{|c|}{unit read 1024, unit write 1024, unit comp \begin{math}1024^4\end{math}} \\
  \hline
  block sizes&8 threads&16 threads&32 threads\\
  \hline
  1&931700&550100&322000\\
  2&718300&626700&388600\\
  4&639200&487700&315700\\
  8&601600&401500&236200\\
  16&560300&351500&197200\\
  32&535800&323600&174000\\
  64&\cellcolor{LightGreen}524200&322700&165800\\
  128&546100&321100&166000\\
  256&632600&\cellcolor{LightGreen}316400&\cellcolor{LightGreen}162900\\
  512&623300&323200&320800\\
  1024&640600&621600&625800\\
  \hline
\end{tabular}
\smallbreak
\textit{AMD 3970X: \begin{math}1024^4\end{math} unit comp}
\medbreak

\endgroup

For the Gold 5225R, 24 threads were run on a single core group, whereas 36 or 48 threads required two core groups; for the AMD 3970X, every four cores form a core group, therefore, the tests listed above covered two, four, and eight core groups.

It is evident that the preferred block size increases by adding core groups, indicating that the task of N iterations would be split into fewer pieces, hence fewer FAA would be triggered. The explanation could be that the cache resynchronizations among the core groups are via media that is markedly less performant than shared L3, such as the hyper-transport link, therefore, FAA overheads are considerably higher than previous cases, accordingly, block sizes should be kept larger to reduce the number of FAA calls.

It has been established that the best block size is proportional to the number of core groups and inversely proportional to the number of threads and size of the unit computation. Further investigation commences on the unit read and write test results:

\smallbreak
\begingroup\centering

\begin{tabular}{ |p{1.6cm}|p{1.6cm}|p{1.6cm}|p{1.6cm}| }
  \hline
  \multicolumn{4}{|c|}{unit read 64, unit write 1024, unit comp \begin{math}1024^6\end{math}} \\
  \hline
  block sizes&4 threads&16 threads&24 threads\\
  \hline
  1&1053600&358800&497400\\
  2&967600&335400&347500\\
  4&937800&252100&280300\\
  8&923400&\cellcolor{LightGreen}245400&247400\\
  16&912400&\cellcolor{LightGreen}244100&240400\\
  32&837600&284700&\cellcolor{LightGreen}153100\\
  64&\cellcolor{LightGreen}511900&270300&197600\\
  128&\cellcolor{LightGreen}549800&436300&202500\\
  256&612900&373100&276500\\
  512&626900&717400&550800\\
  1024&655100&687700&1124600\\
  \hline
\end{tabular}
\smallbreak
\textit{Gold 5225R: 64 unit read}
\medbreak

\begin{tabular}{ |p{1.6cm}|p{1.6cm}|p{1.6cm}|p{1.6cm}| }
  \hline
  \multicolumn{4}{|c|}{unit read 256, unit write 1024, unit comp \begin{math}1024^6\end{math}} \\
  \hline
  block sizes&4 threads&16 threads&24 threads\\
  \hline
  1&736900&352700&299200\\
  2&695500&196200&202100\\
  4&\cellcolor{LightGreen}631600&\cellcolor{LightGreen}209800&\cellcolor{LightGreen}181300\\
  8&\cellcolor{LightGreen}636700&\cellcolor{LightGreen}182000&\cellcolor{LightGreen}178400\\
  16&\cellcolor{LightGreen}634900&\cellcolor{LightGreen}200400&\cellcolor{LightGreen}179100\\
  32&\cellcolor{LightGreen}621700&\cellcolor{LightGreen}187000&192100\\
  64&\cellcolor{LightGreen}632200&238400&261800\\
  128&683600&227500&253500\\
  256&769800&426200&367200\\
  512&790800&480800&781600\\
  1024&787100&654200&1521700\\
  \hline
\end{tabular}
\smallbreak
\textit{Gold 5225R: 256 unit read}
\medbreak

\begin{tabular}{ |p{1.6cm}|p{1.6cm}|p{1.6cm}|p{1.6cm}| }
  \hline
  \multicolumn{4}{|c|}{unit read 4096, unit write 1024, unit comp \begin{math}1024^6\end{math}} \\
  \hline
  block sizes&4 threads&16 threads&24 threads\\
  \hline
  1&\cellcolor{LightGreen}3511000&974500&797600\\
  2&\cellcolor{LightGreen}3436000&948600&766800\\
  4&\cellcolor{LightGreen}3433400&\cellcolor{LightGreen}940300&\cellcolor{LightGreen}767100\\
  8&\cellcolor{LightGreen}3581400&\cellcolor{LightGreen}942900&\cellcolor{LightGreen}742000\\
  16&3841100&975600&794000\\
  32&3845200&942100&808800\\
  64&4071100&1069800&1065700\\
  128&4644200&1110300&1069800\\
  256&4661300&1088300&1760700\\
  512&4638300&2215800&3464700\\
  1024&4212500&3271700&7041200\\
  \hline
\end{tabular}
\smallbreak
\textit{Gold 5225R: 4096 unit read}
\medbreak

\endgroup

From the numbers listed above, the best block size is clearly also inversely proportional to the unit read. Notably, with more engaged threads, the block size decreases accordingly, as was observed in the tests on unit computation. Next, varied unit write statistics are considered:

\smallbreak
\begingroup\centering

\begin{tabular}{ |p{1.6cm}|p{1.6cm}|p{1.6cm}|p{1.6cm}| }
  \hline
  \multicolumn{4}{|c|}{unit read 1024, unit write \begin{math}2^{12}\end{math}, unit comp \begin{math}1024^6\end{math}} \\
  \hline
  block sizes&8 threads&16 threads&32 threads\\
  \hline
  1&1375600&457100&324900\\
  2&1456100&639800&526100\\
  4&1302800&487800&396400\\
  8&1193800&422500&341700\\
  16&\cellcolor{LightGreen}1158400&383800&\cellcolor{LightGreen}309400\\
  32&\cellcolor{LightGreen}1140500&378000&\cellcolor{LightGreen}304800\\
  64&\cellcolor{LightGreen}1127600&\cellcolor{LightGreen}362900&353200\\
  128&1215500&\cellcolor{LightGreen}362800&383700\\
  256&1389300&365300&373100\\
  512&1379700&690000&688800\\
  1024&1384100&1383200&1379600\\
  \hline
\end{tabular}
\smallbreak
\textit{AMD 3970X: \begin{math}2^{12}\end{math} unit write}
\medbreak

\begin{tabular}{ |p{1.6cm}|p{1.6cm}|p{1.6cm}|p{1.6cm}| }
  \hline
  \multicolumn{4}{|c|}{unit read 1024, unit write \begin{math}2^{14}\end{math}, unit comp \begin{math}1024^6\end{math}} \\
  \hline
  block sizes&8 threads&16 threads&32 threads\\
  \hline
  1&3818400&\cellcolor{LightGreen}1259400&\cellcolor{LightGreen}954400\\
  2&3932900&2068000&1273700\\
  4&3758100&2000000&1180800\\
  8&3721500&1755200&1127300\\
  16&\cellcolor{LightGreen}3692200&1931600&1118500\\
  32&\cellcolor{LightGreen}3653800&1641100&1146200\\
  64&\cellcolor{LightGreen}3673800&1577100&1207800\\
  128&3919200&1530900&1316300\\
  256&4445700&1513100&2117700\\
  512&4446400&2774100&2844900\\
  1024&4503600&4421800&4490600\\
  \hline
\end{tabular}
\smallbreak
\textit{AMD 3970X: \begin{math}2^{14}\end{math} unit write}
\medbreak

\begin{tabular}{ |p{1.6cm}|p{1.6cm}|p{1.6cm}|p{1.6cm}| }
  \hline
  \multicolumn{4}{|c|}{unit read 1024, unit write \begin{math}2^{16}\end{math}, unit comp \begin{math}1024^6\end{math}} \\
  \hline
  block sizes&8 threads&16 threads&32 threads\\
  \hline
  1&13781800&\cellcolor{LightGreen}8311600&\cellcolor{LightGreen}7729600\\
  2&\cellcolor{LightGreen}13626800&10221600&11584900\\
  4&\cellcolor{LightGreen}13509000&10178700&11673600\\
  8&\cellcolor{LightGreen}13542500&10186700&11597300\\
  16&13703600&10284900&11546900\\
  32&13749300&10498000&11570900\\
  64&13714900&10178800&11541200\\
  128&14650200&10367800&10099900\\
  256&16670700&9950500&10291700\\
  512&16645700&12600000&12558400\\
  1024&16739400&16789600&16820400\\
  \hline
\end{tabular}
\smallbreak
\textit{AMD 3970X: \begin{math}2^{16}\end{math} unit write}
\medbreak

\endgroup

Predictably, the unit-write tests suggest similar trends to those of unit read; as unit write increases, the preferred block size for better performance decreases. Combining the observations from unit computation as well as read and write, we can assume that, when the task size is ``larger,'' the preferred block size is smaller. Note that the total number of operations in one block should be calculated as follows:

\[complexity\_of\_block = block\_size * task\_size\]
\[task\_size = unit\_read + unit\_write + unit\_comp\]

\section*{Cost model and improvements}

In summary, we have the following observations:

\begin{itemize}
\item The best block size is proportional to the number of core groups (cores share the same L3 in a core group);
\item The best block size is inversely proportional to the number of threads, unit read, unit write, and unit computation.
\end{itemize}

This confirms that the distribution of the preferred block size varies following fixed rules.
Intuitively, the proposed cost model is formulated as follows:

\[B = \dfrac{\alpha * G + \delta_0}{\beta_0 * T + \beta_1 * R + \beta_2 * W + \beta_3 * C + \delta_1} \]

where B is the expected block size; G is the number of core groups; and T, R, W, and C respectively represent the number of threads, unit read, unit write, and unit computation. \textalpha\, \textbeta\, and \textdelta\ are unknown parameters that require tuning.
To determine the parameters and evaluate the model's efficacy, we implemented a linear regression model using Pytorch \cite{Pytorch}.

\begin{lstlisting}[language=Python]

class CostModel(nn.Module):
    def __init__(self):
        super(CostModel, self).__init__()
        self.power = nn.Linear(1,1)
        self.cost = nn.Linear(4,1)

    def forward(self, x):
        power = self.power(x[:,:1])
        cost = self.cost(x[:,1:])
        return torch.div(power, cost)

\end{lstlisting}

where x is the batched input vectors of the core groups, threads, unit read, unit write, and unit computation. The raw training input is the list of previously collected numbers. Each training vector has six columns.

\bigbreak
\begingroup\centering
\begin{tabular}{ |p{0.8cm}|p{1cm}|p{1cm}|p{1cm}|p{1cm}|p{1cm}| }
  \hline
  G&T&R&W&C&B\\
  \hline
  1&2&1024&1024&1024&128\\
  1&2&1024&1024&\(1024^2\)&64\\
  1&2&1024&1024&\(1024^3\)&32\\
  1&2&1024&1024&\(1024^4\)&16\\
  1&2&1024&1024&\(1024^5\)&8\\
  1&2&1024&1024&\(1024^6\)&4\\
  \hline
\end{tabular}
\endgroup
\bigbreak

Note that the last column is the preferred block size, which is excluded from x.
Next, owing to the overly sparsified inputs compared with the expected output, we performed data normalization in case the training converges slowly, albeit with acceptable losses \cite{vandergoot2017normalize}.
Hence, we did the following:

\begin{itemize}
\item Multiple core group with 100
\item Replace units read and write with n, such that \(2^n\) = unit read/write
\item Replace unit computation with p, such that unit computation = \(2^{10p}\)
\end{itemize}

Thereby, we have better-shaped training data:

\bigbreak
\begingroup\centering
\begin{tabular}{ |p{0.8cm}|p{1cm}|p{1cm}|p{1cm}|p{1cm}|p{1cm}| }
  \hline
  G&T&R&W&C&B\\
  \hline
  100&2&10&10&1&128\\
  100&2&10&10&2&64\\
  100&2&10&10&3&32\\
  100&2&10&10&4&16\\
  100&2&10&10&5&8\\
  100&2&10&10&6&4\\
  \hline
\end{tabular}
\endgroup
\bigbreak

Finally, the cost function is:
\[ loss = (y - y')^2 \].

The training was then conducted on an NVidia Quadro M4000 with the Cuda Toolkit 11.4 \cite{cuda}. After 30 h and approximately \begin{math}10^7\end{math} epochs, the training data loss on over 200 cases was reduced to 2001.48. Hence, for each input data, on average, the loss was less than 10. Several examples are listed in the following table:

\bigbreak
\begingroup\centering
\begin{tabular}{ |p{1cm}|p{0.5cm}|p{0.5cm}|p{0.5cm}|p{0.5cm}|p{0.7cm}|p{1.5cm}| }
  \hline
  G&T&R&W&C&B&Inferred B\\
  \hline
  100&2&10&10&1&128&\cellcolor{LightGreen}125\\
  100&2&10&10&3&64&\cellcolor{LightGreen}51\\
  100&2&10&10&4&32&\cellcolor{LightGreen}39\\
  100&2&10&10&6&16&\cellcolor{LightGreen}27\\
  100&8&10&10&2&32&\cellcolor{LightGreen}36\\
  100&8&10&10&3&32&\cellcolor{LightGreen}30\\
  100&8&10&10&5&16&\cellcolor{LightGreen}22\\
  100&8&10&10&5&16&\cellcolor{LightGreen}22\\
  100&4&6&10&6&64&\cellcolor{LightGreen}80\\
  100&4&8&10&6&32&\cellcolor{LightGreen}37\\
  100&4&12&10&6&16&\cellcolor{LightGreen}17\\
  100&4&16&10&6&16&\cellcolor{LightGreen}11\\
  100&8&8&10&6&16&\cellcolor{LightGreen}27\\
  100&8&10&10&6&16&\cellcolor{LightGreen}19\\
  100&8&16&10&6&4&\cellcolor{LightGreen}10\\
  200&8&10&10&1&128&\cellcolor{LightGreen}108\\
  200&8&10&10&2&64&\cellcolor{LightGreen}85\\
  200&8&10&6&6&64&\cellcolor{LightGreen}112\\
  200&8&10&8&6&64&\cellcolor{LightGreen}65\\
  200&8&10&10&6&64&\cellcolor{LightGreen}46\\
  200&8&10&14&6&32&\cellcolor{LightGreen}29\\
  200&8&10&16&6&16&\cellcolor{LightGreen}24\\
  400&16&6&10&6&128&\cellcolor{LightGreen}126\\
  400&16&8&10&6&128&\cellcolor{LightGreen}92\\
  800&32&6&10&6&128&\cellcolor{LightGreen}136\\
  800&32&10&10&6&64&\cellcolor{LightGreen}98\\
  800&32&16&10&6&64&\cellcolor{LightGreen}69\\
  \hline
\end{tabular}
\endgroup
\medbreak

The formula with trained weights thereby becomes:

\[B = \dfrac{1558.31 - 61.84*G}{693.13 -10.48*T - 33.71*R -34.50*W -26.84*C} \]

\section*{Related work and comparison}
A recently published multi-threading library with public access on Github ---Taskflow[7]---has implemented the ParallelFor semantic. According to the corresponding paper, Taskflow provides a powerful interface that assembles tasks in topological order and executes with maximum parallelism. Regarding ParallelFor (under the name of for\_each), atomic FAA synchronized threads compete for the assigned iterations. Each time a thread attempts to acquire a range of [begin, end), it intends to multiply a decimal constant as q (\begin{math}= 0.5/<number of threads>\end{math}) with the unfinished part of N, defined as r. Thereafter, the end is equated to ``begin + q * r,'' implying that the block size is ``q * r.'' Subsequently, when r is smaller than 4 * <number of threads>, the block size will reduce to 1 until the execution ends.
This approach, though somewhat dynamic, is generally less performant than the one we have implemented herein with the cost model. The performance data for comparison are as follows:

\medbreak
\begingroup\centering
\begin{tabular}{ |p{1.5cm}|p{1.5cm}|p{1.5cm}|p{1.7cm}| }
  \hline
  unit\_read&Taskflow&CostModel&block sizes\\
  \hline
  \(2^6\)&3205000&257100&46\\
  \(2^8\)&420400&259500&27\\
  \(2^{10}\)&462600&390400&19\\
  \(2^{12}\)&1364700&1242900&15\\
  \(2^{14}\)&5822300&4470400&12\\
  \(2^{16}\)&19203300&16524300&10\\
  \hline
\end{tabular}
\smallbreak
\textit{W-3225R: 8 T, 1024 unit\_write,\begin{math}2^{60}\end{math} unit\_comp, in clocks}
\endgroup
\medbreak

\begingroup\centering
\begin{tabular}{ |p{1.5cm}|p{1.5cm}|p{1.5cm}|p{1.7cm}| }
  \hline
  unit\_write&Taskflow&CostModel&block sizes\\
  \hline
  \(2^6\)&580100&403400&48\\
  \(2^8\)&673500&377000&28\\
  \(2^{10}\)&1167500&430700&19\\
  \(2^{12}\)&1077100&847600&15\\
  \(2^{14}\)&3707600&3746300&12\\
  \(2^{16}\)&15411800&15498900&10\\
  \hline
\end{tabular}
\smallbreak
\textit{W-3225R: 8 T, 1024 unit\_read,\begin{math}2^{60}\end{math} unit\_comp, in clocks}
\endgroup
\medbreak

\begingroup\centering
\begin{tabular}{ |p{1.5cm}|p{1.5cm}|p{1.5cm}|p{1.7cm}| }
  \hline
  unit\_comp&Taskflow&CostModel&block sizes\\
  \hline
  1024&1334200&750500&46\\
  \(1024^2\)&790100&744800&36\\
  \(1024^3\)&496600&456000&30\\
  \(1024^4\)&508500&412700&25\\
  \(1024^5\)&527300&424900&22\\
  \(1024^6\)&479000&435300&19\\
  \hline
\end{tabular}
\smallbreak
\textit{W-3225R: 8 T, 1024 unit\_read/write, in clocks}
\endgroup
\medbreak

\begingroup\centering
\begin{tabular}{ |p{1.5cm}|p{1.5cm}|p{1.5cm}|p{1.7cm}| }
  \hline
  unit\_read&Taskflow&CostModel&block sizes\\
  \hline
  \(2^6\)&420900&172200&17\\
  \(2^8\)&459800&157100&13\\
  \(2^{10}\)&764700&228400&11\\
  \(2^{12}\)&797400&633500&9\\
  \(2^{14}\)&3514300&3542800&8\\
  \(2^{16}\)&16775400&14511900&7\\
  \hline
\end{tabular}
\smallbreak
\textit{G-5225R: 24 T, 1024 unit\_write,\begin{math}2^{60}\end{math} unit\_comp, in clocks}
\endgroup
\medbreak

\begingroup\centering
\begin{tabular}{ |p{1.5cm}|p{1.5cm}|p{1.5cm}|p{1.7cm}| }
  \hline
  unit\_write&Taskflow&CostModel&block sizes\\
  \hline
  \(2^6\)&1187900&507100&17\\
  \(2^8\)&909300&362500&13\\
  \(2^{10}\)&537700&258000&11\\
  \(2^{12}\)&1031500&593400&9\\
  \(2^{14}\)&4200200&5229200&8\\
  \(2^{16}\)&28460100&28978100&7\\
  \hline
\end{tabular}
\smallbreak
\textit{G-5225R: 24 T, 1024 unit\_read,\begin{math}2^{60}\end{math} unit\_comp, in clocks}
\endgroup
\medbreak

\begingroup\centering
\begin{tabular}{ |p{1.5cm}|p{1.5cm}|p{1.5cm}|p{1.7cm}| }
  \hline
  unit\_comp&Taskflow&CostModel&block sizes\\
  \hline
  1024&604700&287000&17\\
  \(1024^2\)&549700&272300&15\\
  \(1024^3\)&439200&195100&14\\
  \(1024^4\)&211390&192000&13\\
  \(1024^5\)&367100&190100&12\\
  \(1024^6\)&402900&186100&11\\
  \hline
\end{tabular}
\smallbreak
\textit{G-5225R: 24 T, 1024 unit\_read/write, in clocks}
\endgroup
\medbreak

\begingroup\centering
\begin{tabular}{ |p{1.5cm}|p{1.5cm}|p{1.5cm}|p{1.7cm}| }
  \hline
  unit\_read&Taskflow&CostModel&block sizes\\
  \hline
  \(2^6\)&312700&269600&13\\
  \(2^8\)&348200&182100&11\\
  \(2^{10}\)&367600&320000&9\\
  \(2^{12}\)&819500&337500&8\\
  \(2^{14}\)&1913400&1382500&7\\
  \(2^{16}\)&7120100&4541800&6\\
  \hline
\end{tabular}
\smallbreak
\textit{AMD 3970X: 64 T, 1024 unit\_write,\begin{math}2^{60}\end{math} unit\_comp}
\endgroup
\medbreak

\begingroup\centering
\begin{tabular}{ |p{1.5cm}|p{1.5cm}|p{1.5cm}|p{1.7cm}| }
  \hline
  unit\_write&Taskflow&CostModel&block sizes\\
  \hline
  \(2^6\)&354300&199000&13\\
  \(2^8\)&339200&183500&11\\
  \(2^{10}\)&374600&274100&9\\
  \(2^{12}\)&514400&320300&8\\
  \(2^{14}\)&1488600&1569900&7\\
  \(2^{16}\)&8166500&8368800&6\\
  \hline
\end{tabular}
\smallbreak
\textit{AMD 3970X: 64 T, 1024 unit\_read,\begin{math}2^{60}\end{math} unit\_comp}
\endgroup
\medbreak

\begingroup\centering
\begin{tabular}{ |p{1.5cm}|p{1.5cm}|p{1.5cm}|p{1.7cm}| }
  \hline
  unit\_comp&Taskflow&CostModel&block sizes\\
  \hline
  1024&413800&34500&13\\
  \(1024^2\)&439100&315100&12\\
  \(1024^3\)&413300&358900&11\\
  \(1024^4\)&444500&340500&10\\
  \(1024^5\)&5496700&352200&10\\
  \(1024^6\)&398700&336800&9\\
  \hline
\end{tabular}
\smallbreak
\textit{AMD 3970X: 64 T, 1024 unit\_read/write}
\endgroup
\medbreak

As the listed results indicate, with the fine-tuned cost model included in ParallelFor, a general performance boost of over 20\% was observed. There are also multiple cases in which ParallelFor with the cost model required less than one-fifth of the time taken by Taskflow. Admittedly, there are also several cases in which ParallelFor underperforms compared to Taskflow, however, the lag in each case is relatively negligible. Moreover, we believe that these issues could be solved by using fine-tuned training data that are more representative; for example, the training data could be enriched with more fine-grained case information on task size.

\section*{Conclusion and future work}
Based on the analysis and results presented in this paper, we demonstrated that atomic FAA has a significant effect on the end-to-end latency of ParallelFor on various platforms. Furthermore, by rigorously examining the collected performance statistics on neutralizing the FAA overheads, the best block size for splitting tasks within ParallelFor was found to be virtually dependent on a few parameters, such as core groups, the number of threads, and task size. Based on this information, a cost model is proposed. The training and inferencing of the model confirmed that it was effective when applied to the collected statistics and offers a significant advantage over existing counterparts that employ different approaches.

In future work, the CPU frequency and cache latency parameters, which may significantly promote cost model precision on general platforms, must be further investigated.

\printbibliography

\end{document}